# Design and Performance Evaluation of A New Proposed Fittest Job First Dynamic Round Robin (FJFDRR) Scheduling Algorithm

Prof. Rakesh Mohanty[1]
Lecturer
Department of Computer Science and Engineering
Veer Surendra Sai University of Technology, Burla
Sambalpur, Orissa, India
[1]rakesh.iitmphd@gmail.com

Manas Das[2], M. Lakshmi Prasanna[3], Sudhashree[4]
Students
Department of Computer Science and Engineering
Veer Surendra Sai University of Technology, Burla
Sambalpur, Orissa, India
[2]mrdoitcse@gmail.com, [3]lakshmi.vssut@gmail.com,
[4]sudhashree.uce@gmail.com

*Abstract*— In this paper, we have proposed a new variant of Round Robin scheduling algorithm by executing the processes according to the new calculated Fit Factor 'f' and using the concept of dynamic time quantum. We have compared the performance of our proposed Fittest Job First Dynamic Round Robin(FJFDRR) algorithm with the Priority Based Static Round Robin(PBSRR) algorithm. Experimental results show that our proposed algorithm performs better than PBSRR in terms of reducing the number of context switches, average waiting time and average turnaround time.

## I. INTRODUCTION

Scheduling the processes is one of the primary job of an operating system. The main goal of the scheduling is the maximization of CPU utilization, throughput and minimization of response time, waiting time and turnaround time. The scheduling is used in the real time applications like routing of data packets in computer networking, controlling traffic in airways, roadways and railways etc. Existing non-preemptive algorithms use only one of the basic parameters such as arrival time, user priority and burst time. Preemptive Round Robin implements static time quantum. These two drawbacks of existing algorithms motivate us to design a new algorithm which uses more than one basic parameters and dynamic time quantum concept to improve the performance metrics of existing scheduling algorithms.

### A. Well known CPU Scheduling algorithms

The algorithm used by CPU scheduler, which decides the sequence of execution of the processes waiting in the ready queue is called CPU scheduling algorithm. Some well known CPU scheduling algorithms are First Come First Serve (FCFS), Shortest Job First (SJF), Shortest Remaining Time Next(SRTN), Priority and Round Robin(RR) etc. FCFS schedules the processes in order of their arrival in the ready queue. If the longer process executes first, then waiting time for later shorter processes will increase. To minimize the waiting time, SJF favors the shortest processes to be executed first. But longer processes may starve. The Priority algorithm schedules the processes in order of the priority number assigned to each of the processes. Above mentioned algorithms are non-preemptive and not suitable for time sharing systems. Shortest Remaining Time Next(SRTN) and Round Robin(RR) are preemptive in nature. RR is most suitable for time sharing systems as it improves the responsiveness of the system.

### B. Related Work

Elaborate discussion on various well known CPU scheduling algorithms can be found in [1] and [2]. Best Job First proposed by Al-Husain[3] combined the basic functions of non-preemptive scheduling algorithms. The static time quantum which is a limitation of RR was removed by taking dynamic time quantum by Matarneh[4]. The time quantum that was repeatedly adjusted according to the burst time of the running processes is considered to improve the waiting time, turnaround time and number of context switches. Recently a number of new variants of Improved RR algorithms have been developed in [5, 6, 7, 8, 9, 10].

### C. Our Contribution

In our work, we have scheduled the processes giving importance to both the user priority and shortest burst time priority rather than using single parameter. A new Fit factor 'f' is calculated which decides the sequence of the execution of the processes rather the FCFS sequence as generally happens in RR. The limitation of RR algorithm is static time quantum, so we have used the concept of dynamic time quantum. We have compared the performance of our proposed Fittest Job First Dynamic Round Robin(FJFDRR) algorithm with the Priority Based Static Round Robin(PBSRR) algorithm. Experimental results show that our proposed algorithm performs better than PBSRR.

### D. Organization of the paper

In Section II, the pseudo code and illustration of our proposed FJFDRR algorithm is presented. Section III shows the results of experimental analysis of FJFDRR and its comparison with PBSRR. Conclusion and directions for future work is given in Section IV.

## II. OUR PROPOSED ALGORITHM

### A. Uniqueness of our approach

Generally with every process three factors are associated. These factors are user priority, burst time and arrival time. Above factors play an important role to decide in which sequence the processes will be executed. Sorting according to the importance of these factors, user priority comes first, then the burst time and at last the arrival time of the processes. In FCFS, SJF and Priority algorithms, only one among the three factors are taken into consideration. If we mix up all the 3 factors to calculate a new factor i.e. Fit Factor 'f' which will decide the order of execution then average waiting time, average turnaround time and number of context switches will be decreased. But FCFS, SJF and Priority scheduling algorithms are non-preemptive in nature and they can't be used in time sharing systems. So to increase the responsiveness of the system, RR algorithm should be used. Generally in RR algorithm, processes are taken from the ready queue in FCFS manner for execution. But in our algorithm, 'f' is calculated for each process. The process having the lowest 'f' value will be scheduled first. The two important criteria that decides the early execution of processes are – higher user priority and shorter burst time. As user priority has higher importance than other factors, so it is given a weight age of 60% and burst time is given 40%, assuming that all the processes have same arrival time i.e. arrival time = 0. Let the User Priority = UP, User Priority Weight = UW, Shorter Burst time Priority = SP, Burst time Priority Weight = BW. Then Fit Factor 'f' can be calculated as

$$f = UP * UW + SP * BW \quad \text{-------------------- (1)}$$

Dynamic time quantum is used in order to overcome the limitations of static RR. To get the optimal time quantum, median of the remaining burst time is taken as the time quantum.

### B. Pseudo Code for FJFDRR algorithm

Let n be number of processes.
1. Fit Factor 'f' is calculated for each process present in the ready queue according to equation 1.
2. According to the ascending order of the 'f' value, the processes are sorted in the ready queue.
3. While(ready queue != null)
   {
      (a) TQ = median (remaining burst time
                   of all the processes)
      (b) Assign TQ to process $P_i$
          If (i<n) then go to step 3(a)
   }
   End of while
4. Average waiting time, average turnaround time and context switch are calculated
End

In our algorithm all the processes are scheduled using the newly calculated Fit Factor 'f'. The process having the least 'f' value will be scheduled first. Here dynamic time quantum concept is also used to improve the average waiting time, average turnaround time and to decrease the number of context switches.

### C. Illustration

Given the burst sequence 1  35  12  9  98 with user priority 5  2  4  3  1 respectively. The processes were sorted in ascending order of the burst time. The process with least burst time was assigned as highest SP i.e. 1. Here process_id P1 has the highest SP. For each process factor 'f' was calculated using the equation 1. Factors for P1, P2, P3, P4 and P5 were calculated to be 3.4, 2.8, 3.6, 2.6 and 2.6 respectively. In FJFDRR, the processes were scheduled according to the ascending order of the new calculated factor, i.e. the process with least factor 'f' will be scheduled first. If the two processes have the same factor 'f', then they are scheduled according to the user priority. Here P5 was scheduled first, then P4, P2, P1 and P3 respectively. The time slice was assigned as the median of the remaining burst time in FJFDRR. In the first round time slice was calculated to be 12 and was assigned as the time quantum for all the processes. After the first round only P2 and P5 were left in the ready queue with remaining burst time 86 and 23. So in the second round, the time slice was the floor value of the median of the remaining burst time i.e. 54. In the third round as the only remaining process was P5, the remaining burst time 32 was given as the time quantum so that it completes its execution without any context switching. The above process was continued till all the processes were deleted from the ready queue.

## III. EXPERIMENTAL RESULTS

### A. Assumptions

In a uni-processor environment, all the experiments are performed and all the processes are independent. Time slice is assumed to be not more than the maximum burst time. The attributes like burst time, number of processes and the user-priorities of all the processes are known before submitting the processes to the processor. All processes are CPU bound. No processes are I/O bound.

### B. Experimental Frame Work

Taking various inputs and output parameters we have performed many experiments. The input parameters consist of the number of processes, burst time and user-priorities. The output parameters consist of average waiting time, average turnaround time and number of context switches.

### C. Data set

We have performed six experiments for evaluating performance of our new proposed algorithm. For the experiments, we have considered the data set as the processes with burst time in increasing, decreasing and random order respectively. In the above cases, the arrival time was assumed to be the same.

### D. Performance Metrics

We have used three performance metrics for our experimental analysis. *Turn Around Time(TAT):* For the better performance of the algorithm, average turnaround time should be less. *Waiting Time(WT):* For the better performance of the algorithm, average waiting time should be less. *Number of Context Switches(CS):* For the better performance of the algorithm, the number of context switches should be less.

### E. Experiments Performed

To evaluate the performance of our proposed algorithm, we have taken a set of five and eight processes in six different cases. The algorithm works effectively even if it used with a very large number of processes. In each case, we have compared the experimental results of our proposed algorithm with the priority based RR scheduling algorithm with fixed time quantum Q. Here we have assumed a static time quantum Q equal to 15 in priority based RR algorithm.

**Case 1:** We Assume five processes arriving at time = 0, with increasing burst time (P1 = 9, P2 = 15, P 3 = 27, P4 = 43, p5= 82) as shown in Table-I. The Table-II shows the output using Priority based RR algorithm and our new proposed FJFDRR algorithm. Fig. 1 and Fig. 2 show Gantt chart for both the algorithms respectively. Fig. 3 shows the comparison of AWT, TAT, and No. of context switches (CS) taking PBSRR and FJFDRR.

TABLE - I

| Processes | Arrival Time | Burst Time | User priority |
|---|---|---|---|
| P1 | 0 | 9 | 5 |
| P2 | 0 | 15 | 2 |
| P3 | 0 | 27 | 4 |
| P4 | 0 | 43 | 1 |
| P5 | 0 | 82 | 3 |

TABLE - II

| Algorithm | Time Quantum | Avg TAT | Avg WT | CS |
|---|---|---|---|---|
| PBSRR | 15 | 102 | 66.8 | 12 |
| FJFDRR | 27,35,20 | 88 | 53 | 7 |

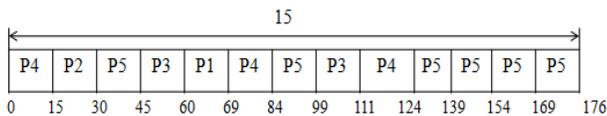

FIG.1: GANTT CHART PBSRR (CASE 1)

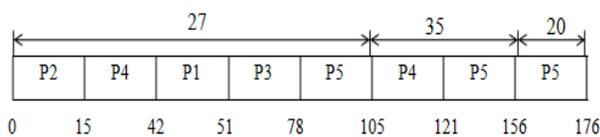

FIG. 2: GANTT CHART FOR FJFDRR (CASE 1)

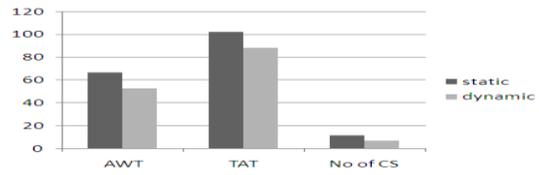

**FIG. 3 : COMPARISON OF PBSRR( Static) and FJFDRR( dynamic) (CASE 1)**

**Case 2:** We Assume eight processes arriving at time = 0, with increasing burst time (P1 = 7, P2 = 20, P 3 = 36, P4 = 53, p5= 69, p6= 82, p7= 94, p8= 100) as shown in Table-III with user priorities. The Table-IV shows the output using PBSRR algorithm and our new proposed FJFDRR algorithm. Fig. 4 and Fig. 5 show Gantt chart for both the algorithms respectively. Fig. 6 shows the comparison of AWT, TAT and No. of context switches (CS) taking static and dynamic time quantum.

TABLE - III

| Processes | Arrival Time | Burst Time | User priority |
|---|---|---|---|
| P1 | 0 | 7 | 8 |
| P2 | 0 | 20 | 1 |
| P3 | 0 | 36 | 6 |
| P4 | 0 | 53 | 3 |
| P5 | 0 | 69 | 2 |
| P6 | 0 | 82 | 5 |
| P7 | 0 | 94 | 4 |
| P8 | 0 | 100 | 7 |

TABLE - IV

| Algorithm | Time Quantum | Avg TAT | Avg WT | CS |
|---|---|---|---|---|
| PBSRR | 15 | 315.25 | 257.62 | 34 |
| FJFDRR | 61, 27, 9, 3 | 282 | 189.5 | 14 |

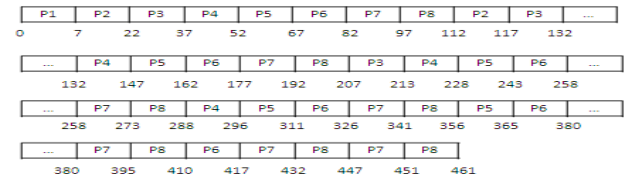

**FIG. 4: GANTT CHART PBSRR (CASE 2)**

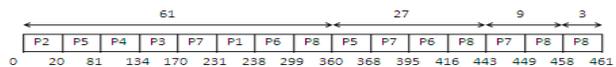

**FIG. 5: GANTT CHART FOR FJFDRR (CASE 2)**

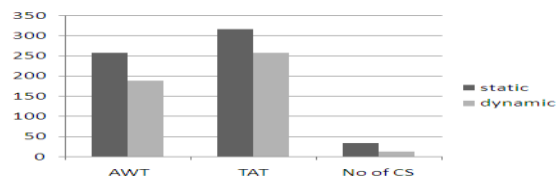

**FIG. 6 : COMPARISON OF PBSRR( Static) and FJFDRR( dynamic) (CASE 2 )**

**Case 3:** We assume five processes arriving at time = 0, with decreasing burst time (P1 = 100, P2 =88, P 3 = 64, P4 = 37, p5= 3) as shown in Table-V. The Table-VI shows the output using PBSRR algorithm and our new proposed FJFDRR algorithm. Fig.7 and Fig. 8 show Gantt chart for both the algorithms respectively. Fig. 9 shows the comparison of AWT, TAT and No. of context switches (CS) taking static and dynamic time quantum.

TABLE - V

| Processes | Arrival Time | Burst Time | User priority |
|---|---|---|---|
| P1 | 0 | 100 | 5 |
| P2 | 0 | 88 | 3 |
| P3 | 0 | 64 | 1 |
| P4 | 0 | 37 | 4 |
| P5 | 0 | 3 | 2 |

TABLE - VI

| Algorithm | Time Quantum | Avg TAT | Avg WT | CS |
|---|---|---|---|---|
| PBSRR | 15 | 192.19 | 133.8 | 21 |
| FJFDRR | 64, 30, 6 | 144.4 | 86 | 7 |

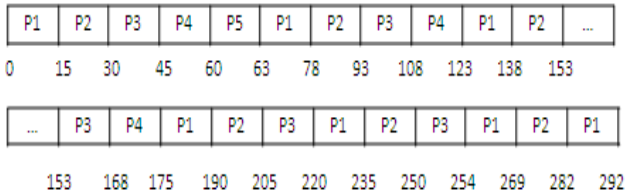

FIG. 7: GANTT CHART PBSRR (CASE 3)

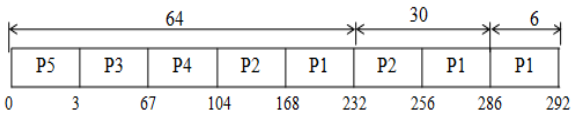

FIG. 8: GANTT CHART FOR FJFDRR (CASE 3)

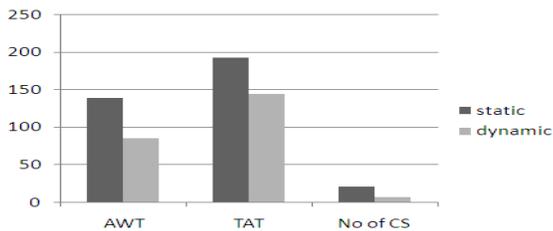

FIG. 9 : COMPARISON OF PBSRR( Static) and FJFDRR( dynamic) (CASE 3 )

**Case 4:** We assume five processes arriving at time = 0, with random burst time (P1 = 1, P2 =35, P 3 = 12, P 4 = 9, p5= 98) as shown in Table-VII. The Table-VIII shows the output using PBSRR algorithm and our new proposed FJFDRR algorithm. Fig. 10 and Fig. 11 show Gantt chart for both the algorithms respectively. Fig. 12 shows the comparison of AWT, TAT and No. of context switches (CS) taking static and dynamic time quantum.

TABLE-VII

| Processes | Arrival Time | Burst Time | User priority |
|---|---|---|---|
| P1 | 0 | 1 | 5 |
| P2 | 0 | 35 | 2 |
| P3 | 0 | 12 | 4 |
| P4 | 0 | 9 | 3 |
| P5 | 0 | 98 | 1 |

TABLE-VIII

| Algorithm | Time Quantum | Avg TAT | Avg WT | CS |
|---|---|---|---|---|
| PBSRR | 15 | 79.8 | 48.8 | 12 |
| FJFDRR | 12, 54, 32 | 75.8 | 44.8 | 7 |

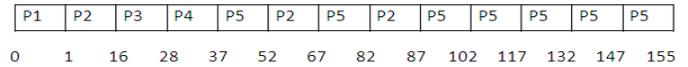

FIG. 10: GANTT CHART PBSRR (CASE 4)

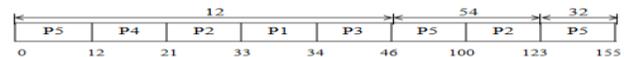

FIG. 11: GANTT CHART FJFDRR (CASE 4)

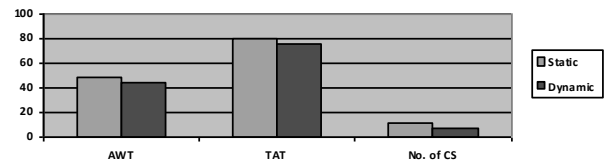

FIG. 12 : COMPARISON OF PBSRR( Static) and FJFDRR( dynamic) (CASE 4 )

**Case 5**: We Assume eight processes arriving at time = 0, with random burst time (P1 = 25, P2 = 99, P 3 = 9, P4 = 32, p5= 68, p6=75, p7= 17, p8= 2) as shown in Table- IX with user priorities. The Table- X shows the output using PBSRR algorithm and FJFDRR. Fig. 13 and Fig. 14 show Gantt chart for both the algorithms respectively. Fig. 15 shows the comparison of AWT, TAT and No. of context switches (CS) taking static and dynamic time quantum.

TABLE - IX

| Processes | Arrival Time | Burst Time | User priority |
|---|---|---|---|
| P1 | 0 | 25 | 3 |
| P2 | 0 | 99 | 6 |
| P3 | 0 | 9 | 7 |
| P4 | 0 | 32 | 1 |
| P5 | 0 | 68 | 8 |
| P6 | 0 | 75 | 5 |
| P7 | 0 | 17 | 2 |
| P8 | 0 | 2 | 4 |

TABLE - X

| Algorithm | Time Quantum | Avg TAT | Avg WT | CS |
|---|---|---|---|---|
| PBSRR | 15 | 183 | 142.13 | 25 |
| FJFDRR | 28, 43, 16, 12 | 164.5 | 122.13 | 14 |

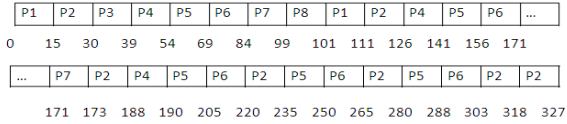

FIG. 10: GANTT CHART PBSRR (CASE 5)

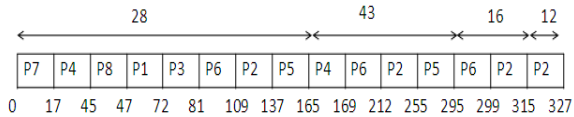

FIG. 10: GANTT CHART FJFDRR (CASE 5)

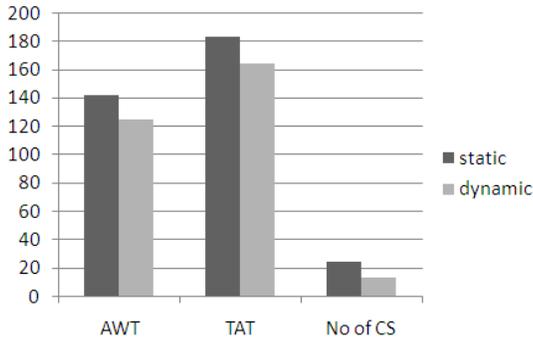

FIG. 12 : COMPARISON OF PBSRR( Static) and FJFDRR( dynamic) (CASE 5 )

## IV. CONCLUSION

From the experimental results, we found that FJFDRR performs better than the PBSRR in terms of decreasing the number of context switches, average waiting time and average turn around time. Implementing the arrival time factor in FJFDRR can be done in future research work.

AUTHORS PROFILE

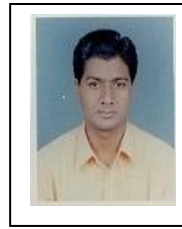

**Prof Rakesh Mohanty** is currently a lecturer in Department of Computer Science and Engineering, Veer Surendra Sai University of Technology, Burla, Orissa, India. He received his B. E. degree in Computer Science and Engineering from University College of Engineering(UCE), Burla in 1998 and M. Tech. degree in Computer Science and Technology from School of Computer and Systems Sciences, Jawaharlal Nehru University(JNU), Newdelhi, India in 2002. He is pursuing his PhD in Computer Science and Engineering at Indian Institute of Technology(IIT), Madras, India under the Supervision of Dr. N. S. Narayanaswamy in the area of Design and Analysis of Algorithms. His broad research area of interest includes Algorithms, Data Structures, Scheduling and Paging. He has published 15 research papers in various Refered International Journals and International Conference Proceedings. He has more than 10 years of teaching experience. He is a scholarship and top rank holder through out his academic career. He is a member of International Association of Engineers(MIAENG).

Manas Das, M. Lakshmi Prasanna, Sudhashree have received their B. Tech. in Computer Science and Engineering from Veer Surendra Sai University of Technology, Burla, Sambalpur, Orissa, India in 2010.